\documentclass[prl,aps,twocolumn]{revtex4-1}

\usepackage[utf8]{inputenc}

\usepackage{amsmath,amssymb,latexsym,epsfig,graphics,epsf}
\usepackage{hyperref}

\usepackage{color}
\usepackage{dcolumn}

\newcommand{\be}{\begin{equation}}
\newcommand{\ee}{\end{equation}}

\begin{document}
	\renewcommand{\i}{\operatorname{i}}
	
	\title{Solid-liquid coexistence of the noble elements. II. Neon, krypton and xenon}
	\author{Aditya N. Singh}
	\affiliation{Theoretical Chemistry Institute and Department of Chemistry, University of Wisconsin-Madison, 1101 University Avenue, Madison, Wisconsin 53703, USA}
	\author{Jeppe C. Dyre}
	\author{Ulf R. Pedersen}
	\email{ulf@urp.dk}
	\affiliation{{\it Glass and Time}, IMFUFA, Department of Science and Environment, Roskilde University, P. O. Box 260, DK-4000 Roskilde, Denmark}
	
	\date{\today}
	
	\begin{abstract}
      The noble elements constitute the simplest group in the periodic table. At low temperatures or high pressures, the liquid phase solidifies into a face-centered cubic crystal structure (except helium). In the companion paper, we investigated the fcc solid-liquid coexistence of argon in the light of hidden scale invariance. Here we extend the investigation to neon, argon, krypton, and xenon. Computations are done using the SAAP potential, suggested by Deiters and Sadus [J. Chem. Phys {\bf 150}, 134504 (2019)], derived from accurate {\it ab initio} calculations. The systems exhibit hidden scale invariance in the investigated part of the phase diagram, which makes it possible to predict the shape and property variations along the solid-liquid coexistence lines.
	\end{abstract}
	
	\maketitle
	
	\section{Introduction}
	This paper we investigating the solid-liquid coexistence of noble elements. The companion paper, Ref.\ \cite{paperI}, presented results for the argon (Ar) parametrization of the simplified ab initio
	atomic (SAAP) potential recently suggested by Deiters and Sadus \cite{deiters2019}. Here, we extend the investigation to include the noble elements Ne, Kr, and Xe. The parameters for the SAAP potential are determined from {\it ab initio} quantum-mechanical calculations using the coupled-cluster approach \cite{konrad2010,bartlett2007} on the CCSD(T) theoretical level \cite{nasrabad2004, smits2020}. 
	
\begin{figure}
		\begin{center}
			\includegraphics[width=0.49\textwidth]{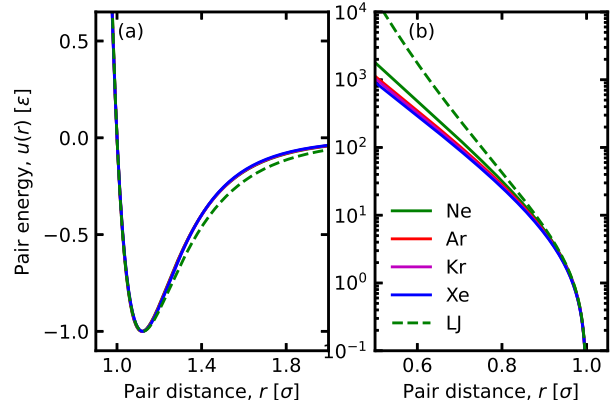}
		\end{center}
		\caption{\label{fig:SAAP} The SAAP pair potential $v(r)$, Eq.\ (\ref{eq:saap}), of Ne (green solid), Ar (red solid), Ke (purple solid), Xe (blue solid), and LJ (green dashed) on (a) a linear and (b) a logarithmic scale.}
	\end{figure}

	We perform molecular dynamics as presented in paper I \cite{paperI}: We consider monatomic systems of $N$ particles with mass $m$ confined in a volume $V$ with periodic boundaries for the number density $\rho=N/V$. Let ${\bf R} = ( {\bf r}_1, {\bf r}_2, {\bf r}_3,\ldots,{\bf r}_N)$ be the collective coordinate vector. The potential-energy function is a sum of pair potential contributions,
	$	U({\bf R}) = \sum_{i>j}^N\varepsilon\, v(|{\bf r}_i-{\bf r}_j|/\sigma)
$	where the SAAP pair potential is
	\begin{equation}\label{eq:saap}
	v(r) = \frac{a_0\exp(a_1r)/r+a_2\exp(a_3r)+a_4}{1+a_5r^6}.
	\end{equation}
	For each of the noble elements the six $a_i$ parameters can be found in Reference \cite{deiters2019}; they are determined by fitting to results of the above mentioned {\it ab initio} calculations on dimers \cite{konrad2010}. The SAAP pair potential is truncated and shifted at $r_c=4$ in units of $\sigma$.
	Figure \ref{fig:SAAP}(a) shows the pair potentials of Ne, Ar, Kr, Xe in units of $\sigma$ and $\varepsilon$ (Table \ref{table:interface_pinning}) and, for comparison, of the Lennard-Jones (LJ) potential truncated and shifted at $r_c=6$. The SAAP pair potentials are parameterized to have the same minimum in MD units as the LJ pair potential. Note that the Ne pair potential appears to be quite ``hard'' while the Xe pair potential is more ``soft'' at short distances. This difference is reflected in the shapes of the solid-liquid coexistence lines, as shown below.
	
	We use the RUMD software package \cite{rumd} to study systems of $N=5120$ particles in an elongated orthorhombic simulation cell where the box lengths in the $y$ and $z$ directions are equal, and the box length in the $x$ direction is 2.5 times longer. We performed molecular dynamics simulations for $2^{22}\simeq4\times10^6$ steps after equilibration (also $2^{22}$ steps) using a leap-frog time-step of $0.004\sigma\sqrt{m/\varepsilon}$. This result in a simulation time of roughly  $1.7\times10^4\sigma\sqrt{m/\varepsilon}$, corresponding to 33 ns in argon units. The temperature $T$ and pressure $p$ is kept constant using the Langevin type dynamics suggested by Grønbech-Jensen et al.\ \cite{gronbech2014}. 
	
	\section{The Solid-liquid coexistence lines}
\begin{figure*}
		\begin{center}
			\includegraphics[width=0.49\textwidth]{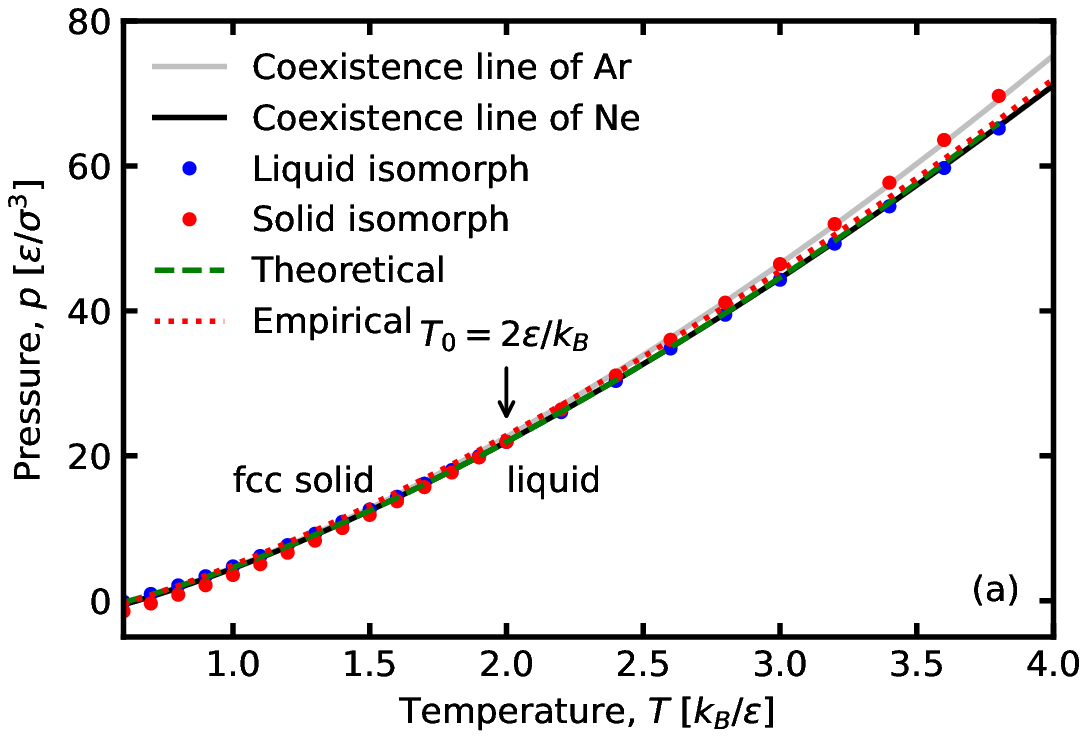}
			\includegraphics[width=0.49\textwidth]{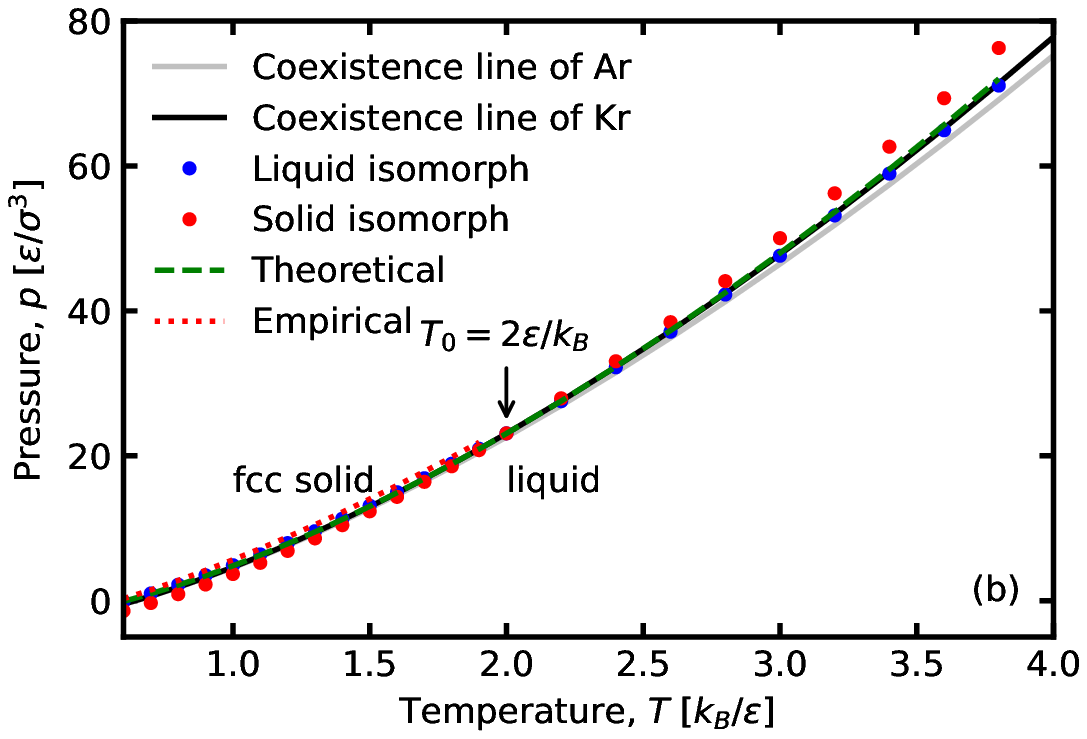}
			\includegraphics[width=0.49\textwidth]{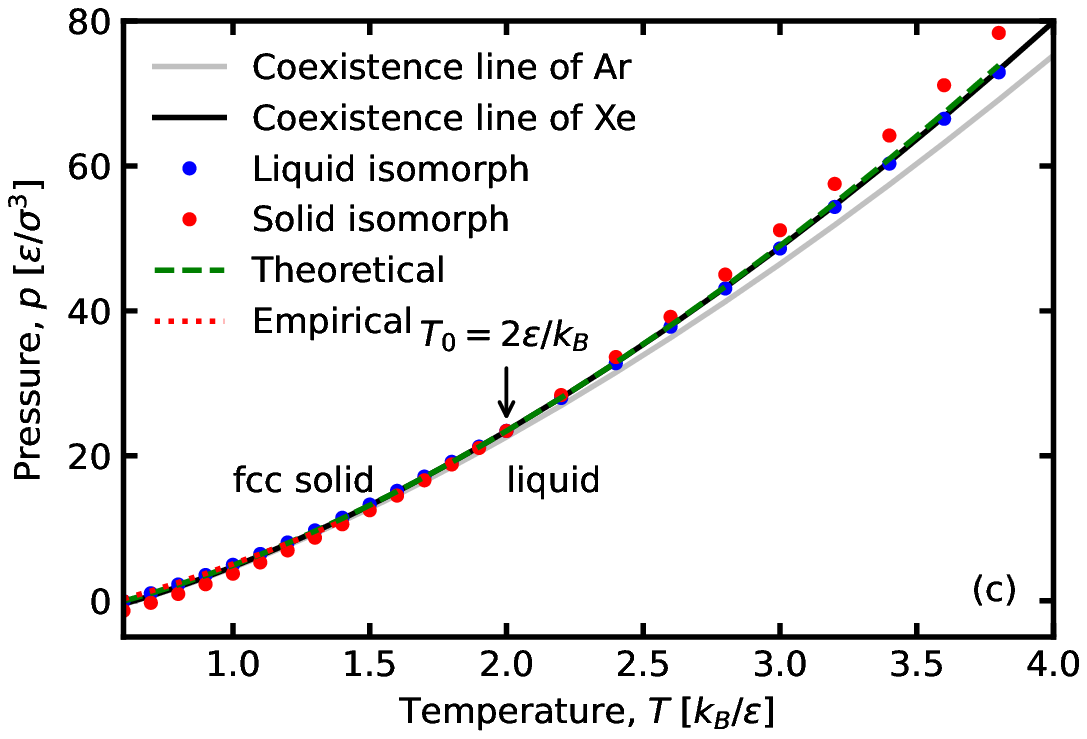}
			\includegraphics[width=0.49\textwidth]{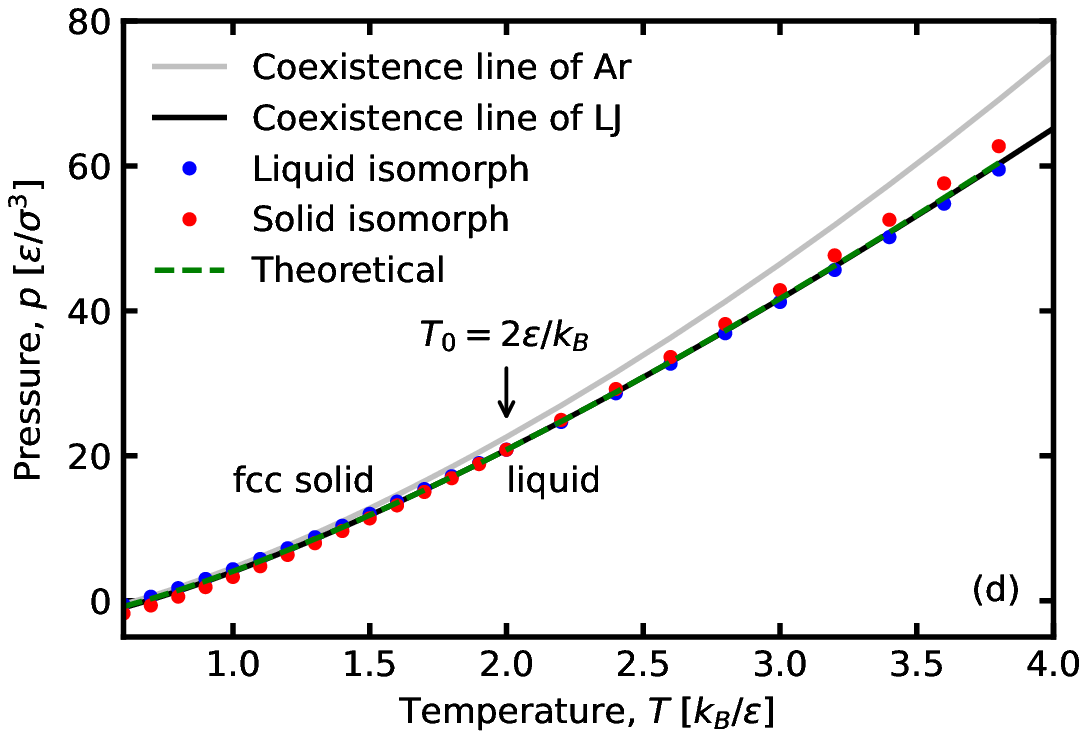}
		\end{center}
		\caption{\label{fig:pressure_temperature} The solid-liquid coexistence line in the pressure-temperature plane of (a) neon, (b) krypton, (c) xenon and (d) the LJ model. For reference, each figure shows the coexistence line of argon (gray line) \cite{paperI}. The dots represent solid and liquid isomorphic state points (generated from the reference state point at $T_0=2\varepsilon/k_B$), the green dashed line is the theoretical prediction of the isomorph theory (see below). Empirical melting lines are shown as red dotted lines \cite{vos1991, ferreira2008}.}
	\end{figure*}

	The solid-liquid coexistence lines of Ne, Kr, and Xe are computed as for Ar in Paper I \cite{paperI}: The interface pinning method \cite{pedersen2013} is used to compute the coexistence state-point at the reference temperature $T_0=2\varepsilon/k_B$. Table \ref{table:interface_pinning} shows the estimated coexistence pressures $p$, volume per particle in the liquid state $V_l/N$, volume per particle in the solid-state $V_s/N$, the volume change per particle at melting $\Delta V_m/N$, and the entropy of melting per particle $\Delta S_m/N$. 
	Other coexistence points are subsequently determined by numerical integration of Clausius-Clapeyron identity, $dP/dT=\Delta S_m/\Delta V_m$, using the standard fourth-order Runge-Kutta algorithm. The slope $dP/dT$ is evaluated from thermodynamic information derived from $NpT$ simulations. The numerical integration is carried out with temperature as the independent variable, starting from the reference temperature $T_0=2\varepsilon/k_B$ and moving down to temperature $0.65\varepsilon/k_B$ and up to $4\varepsilon/k_B$.
	Figures \ref{fig:pressure_temperature}(a)-(c) show the resulting coexistence lines for Ne, Kr and Xe, respectively. For reference, the gray lines on each panel show the Ar coexistence line.
	The dotted red lines are empirical coexistence lines \cite{vos1991,ferreira2008}. The SAAP potential systematically overestimates the coexistence temperature at a given temperature. This is likely due to missing many body interactions, as discussed for the case of Argon in Ref. \cite{paperI}.
	
	\begin{table*}
		\caption{\label{table:interface_pinning} Thermodynamic data for estimated coexistence state points at the (reference) temperature $T_0=2\varepsilon/k_B$ obtained by the interface-pinning method \cite{pedersen2013}.}
		\begin{ruledtabular}
			\begin{tabular}{rll|lllll}
				& $\varepsilon/k_B$ [K] & $\sigma$ [Å] & $p_0$ [$\varepsilon/\sigma^{3}$] & $V_l/N$ [$\sigma^{3}$] & $V_s/N$ [$\sigma^{3}$] & $\Delta V_m/N$ [$\sigma^{3}$] & $\Delta S_m/N$ [$k_B$] \\
				\hline
				Ne & \ 42.36 & 2.759 & 21.911(2) & 0.93482(2) & 0.88069(2) & 0.054128(8) & 1.09707(13) \\
				Ar & 143.5 & 3.355 & 22.591(4) & 0.92612(3) & 0.87489(3) & 0.051227(11) & 1.08515(13) \\
				Kr & 201.1 & 3.580 & 23.079(2) & 0.91999(2) & 0.87075(2) & 0.0492320(9) & 1.07343(16) \\
				Xe & 280.2 & 3.901 & 23.423(4) & 0.91623(3) & 0.86828(3) & 0.047951(7) & 1.06753(11)\\
				LJ & & & 20.8270(8)& 0.940160(8)& 0.882777(7) & 0.057388(7) & 1.09727(13) \\
			\end{tabular}
		\end{ruledtabular}
	\end{table*}
	
	Deiters and Sadus investigated the gas-liquid coexistence lines for the SAAP potentials of the noble elements  \cite{deiters2019b}. Here we compute the coexistence between the liquid and the face-centred cubic (fcc) solid. In all, this information allows us to compute the gas-liquid-fcc triple points (Table \ref{table:triple_points}). The triple point temperatures of the elements are very similar, ranging from $0.64251(35)\varepsilon/k_B$ for Xe to $0.66054(5)\varepsilon/k_B$ for Ne. This is not surprising, given the similar shapes of the pair potentials (Fig.\ \ref{fig:SAAP}). The pair-potential parameters were chosen to have the same minimum in reduced units as the LJ model. However, the triple point temperature of the LJ model is somewhat higher, 0.694$\varepsilon/k_B$ \cite{mastny2007}, which we interpret as an effect of the broader range of attraction of the LJ pair potential compared to that of the SAAP potentials.
	
	\begin{table}
		\caption{\label{table:triple_points} Thermodynamic data for the estimated triple points.}
		\begin{ruledtabular}
			\begin{tabular}{r|lllll}
				& $T_\textrm{tp}$ [$\varepsilon/k_B$] & $\rho_l$ [$\sigma^{-3}$] & $\rho_s$ [$\sigma^{-3}$] & $\rho_g$ [$\sigma^{-3}$] \\
				\hline
				Ne & 0.65054(5) & 0.95239(5) & 0.82592(16) & 0.00421(25) \\
				Ar & 0.64679(34) & 0.94953(5) & 0.82532(35) & 0.00425(25) \\
				Kr & 0.64577(54) & 0.94750(2) & 0.82624(26) & 0.00351(28) \\
				Xe & 0.64251(35) & 0.94534(4) & 0.82378(15) & 0.00438(11) \\
				LJ & 0.694$^1$ & 0.96 & 0.84 \\
			\end{tabular}
		\end{ruledtabular}
		$^1$: The LJ values are from Reference \cite{mastny2007}.
	\end{table} 

\begin{figure}
	\begin{center}
		\includegraphics[width=0.49\textwidth]{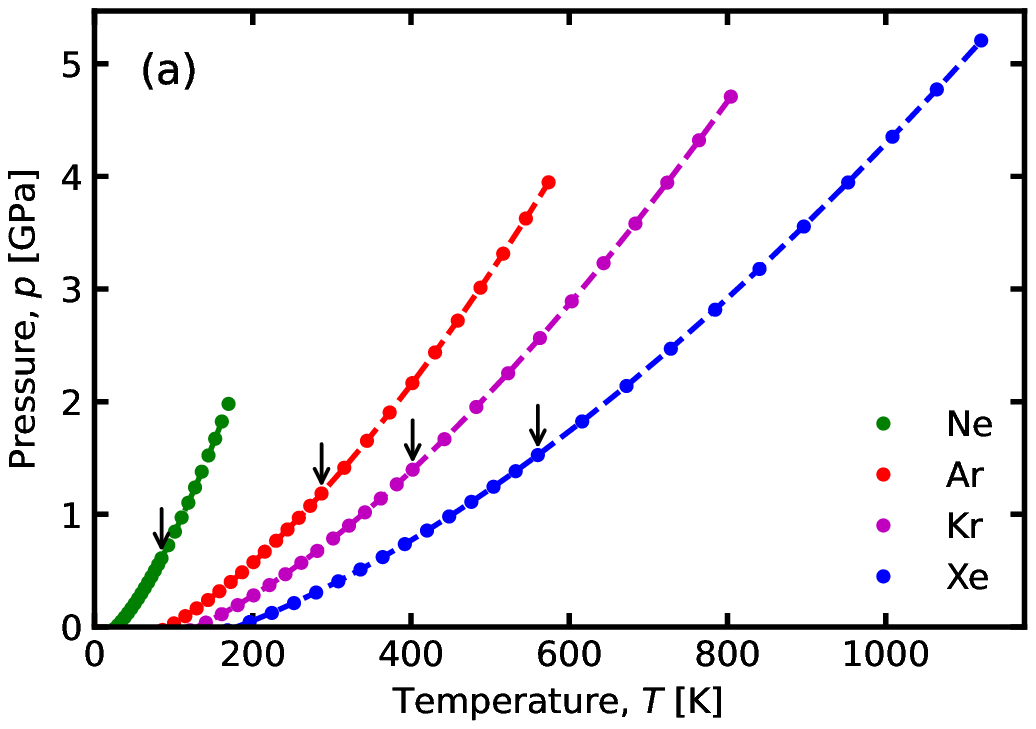}
		\includegraphics[width=0.49\textwidth]{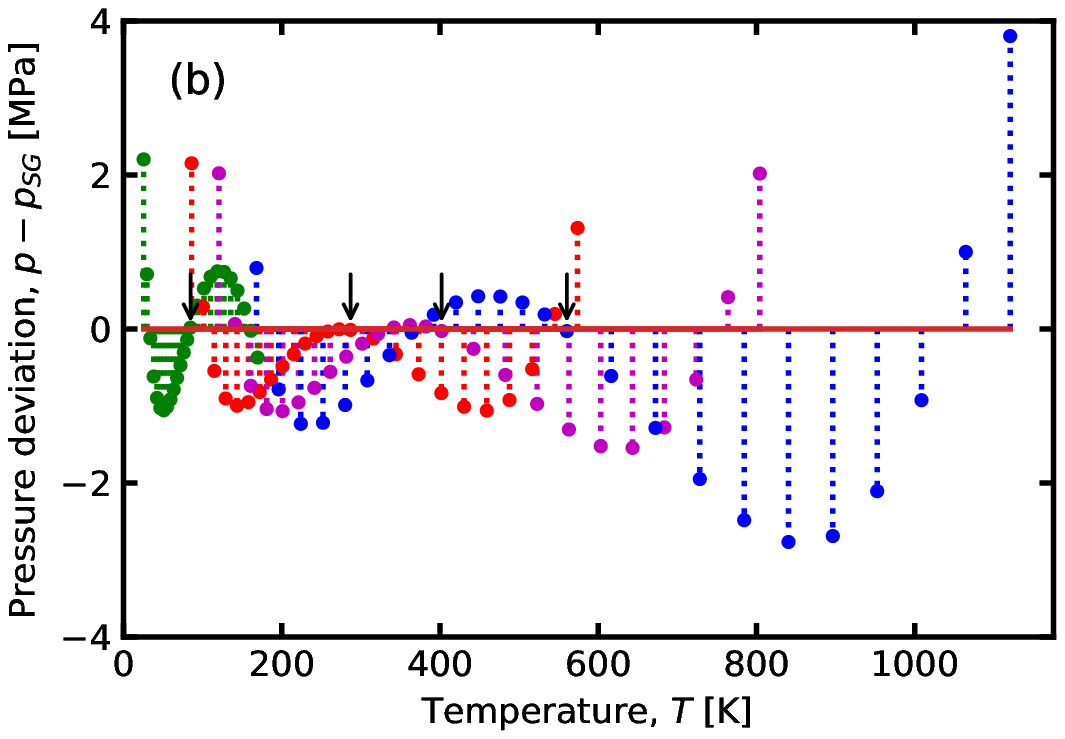}
		\includegraphics[width=0.49\textwidth]{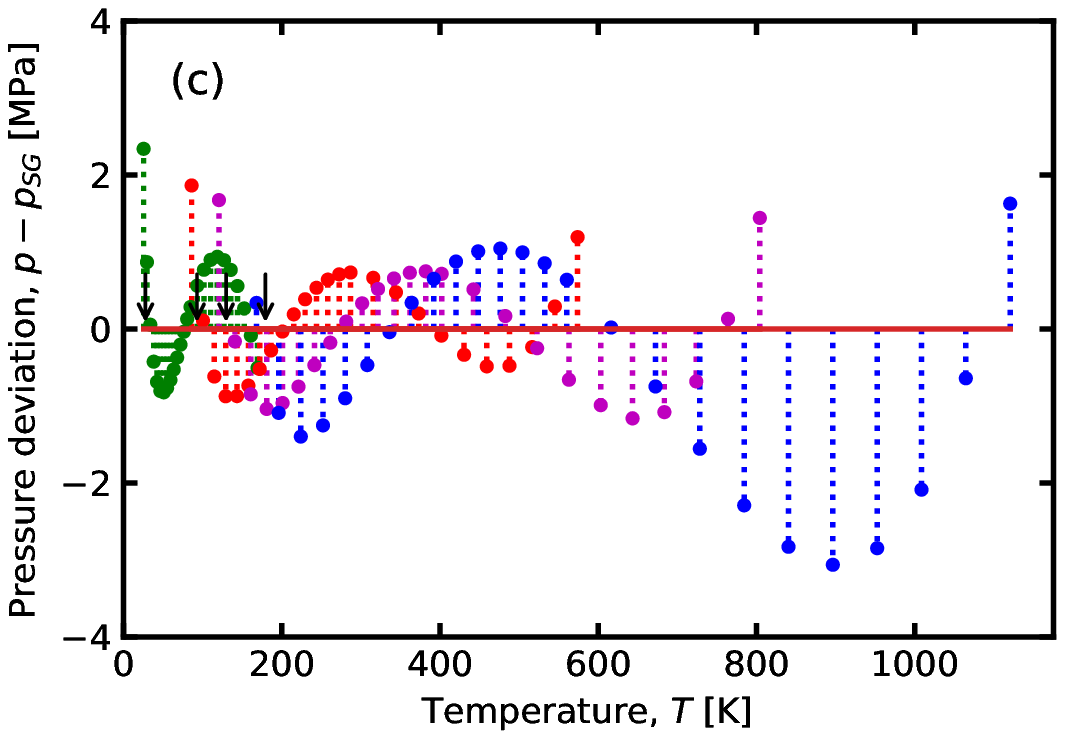}
	\end{center}
	\caption{\label{fig:simon_fit} (a) Fit of the Simon–Glatzel equation (Eq. \ref{eq:simon_fit}; dashed lines) to the SAAP solid-liquid coexistence line (dots) of Ne (green), Ar (red), Kr (purple) and Xe (blue). The reference points ($T_0$, $p_0$) are indicated with arrows. Parameters for Eq. \ref{eq:simon_fit} are given in Table \ref{table:simon}. (b) Residuals using $T_0$ as the reference temperature (indicated with arrows). (c) Residuals using $p_\textrm{ref}=0$ and with the reference temperature $T_\textrm{ref}$ as a fitting parameter ($T_\textrm{ref}\simeq T_\textrm{tp}$ is indicated with arrows). Parameters for Eq. \ref{eq:simon_fit} are given in Table \ref{table:simon_tp}.}
\end{figure}

As an aside, we investigate the validity of the Simon–Glatzel equation for the coexistence pressure \cite{simon1929},
\begin{equation}\label{eq:simon_fit}
p_\textrm{SG}(T) = p_\textrm{ref} + a[[T/T_\textrm{ref}]^c-1].
\end{equation}
We first use $T_\textrm{ref}=T_0$ and $p_\textrm{ref}=p_0$. 
Figure \ref{fig:simon_fit}(a) show fit to the SAAP coexistence lines where the $a$ and $c$ parameters are determined by the least square method (see Table \ref{table:simon}). The accuracy of the fit is within a few MPa (Fig.\ \ref{fig:simon_fit}(b) show residuals). The triple point temperature is often used when fitting empirical data. 
Table \ref{table:simon_tp} gives parameters using $p_\textrm{ref}=0$ and $T_\textrm{ref}$ as a third fitting parameter (in addition to $a$ and $c$). The accuracy of the fit is compatible for the two approaches (see Figs.\ \ref{fig:simon_fit}(b) and \ref{fig:simon_fit}(c)). With the latter procedure, the reference temperature almost identical to the triple point temperature: $T_\textrm{ref}\simeq T_\textrm{tp}$ (since the triple point pressure is nearly zero for the relevant pressure scale). Table \ref{table:simon_tp} compare SAAP parameters with parameters from empirical data. The agreement is in good. The $a$ parameter and $T_\textrm{ref}$ of the SAAP fit is systematically lower than the parameters determined from empirical data. This is likely due to missing many body interactions of the SAAP potential, as discussed for the case of Ar in Ref.\ \cite{paperI}.
In the remainder of the paper we do not use the Simon–Glatzel approximate.

\begin{table}
	\caption{\label{table:simon} Parameters for the Simon–Glatzel equation (Eq.\ \ref{eq:simon_fit}) of SAAP coexistence state points using $T_\textrm{ref}=T_0$ and $p_\textrm{ref}=p_0$ as reference state-point. The fit is shown on and Fig.\ \ref{fig:simon_fit}(a).}
	\begin{ruledtabular}
		\begin{tabular}{rrr|rr}
			& $T_0$ [K] & $p_0$ [GPa] & $a$ [GPa] & c  \\
			\hline
			Ne & 84.72 & 0.61010 & 0.751 & 1.4983 \\
			Ar & 286.98 & 1.18498 & 1.436 & 1.5473 \\
			Kr & 402.16 & 1.39645 & 1.681 & 1.5699 \\
			Xe & 560.37 & 1.52607 & 1.824 & 1.5924 \\
		\end{tabular}
	\end{ruledtabular}
\end{table}

\begin{table}
	\caption{\label{table:simon_tp} Parameters for the Simon–Glatzel equation (Eq.\ \ref{eq:simon_fit} ) of SAAP coexistence state points and of empirical data using the triple point temperature as reference: $T_\textrm{ref}=T_\textrm{tp}$. For the SAAP results, $T_\textrm{ref}$ is treated as a fitting parameter and $p_\textrm{ref}=0$.}
	\begin{ruledtabular}
		\begin{tabular}{rlr|rrr}
			& & $T_\textrm{ref}$ [K] & $a$ [GPa] & $c$ \\
			\hline
			Ne & SAAP & 27.75 & 0.1409 & 1.4989 \\
			   & empirical, Ref.\ \cite{vos1991} & 24.55  & 0.1286 & 1.4587 \\
			Ar & SAAP & 92.91 & 0.2501 & 1.5487 \\
			   & empirical, Ref.\ \cite{ferreira2008} & 83.81 & 0.2245 & 1.5354 \\
			Kr & SAAP & 129.64 & 0.2835 & 1.5713 \\
			  & empirical, Ref.\ \cite{ferreira2008} &  115.77 & 0.2666 & 1.4951 \\			
			Xe & SAAP & 179.45 & 0.2966 & 1.5942 \\
			  & empirical, Ref.\ \cite{ferreira2008} &   161.40   &    0.2594 & 1.4905			
		\end{tabular}
	\end{ruledtabular}
\end{table}

	\section{Hidden scale invariance}
	In the companion paper \cite{paperI} we show that the Ar parameterization of the SAAP potential has hidden scale invariance (for the investigated state-points). This fact was used to make an accurate prediction of the shape of the coexistence line as well as of the variation of several properties along the coexistence line. Below, we apply this theoretical framework to Ne, Kr and Xe.
	
\begin{figure*}
		\begin{center}
			\includegraphics[width=0.49\textwidth]{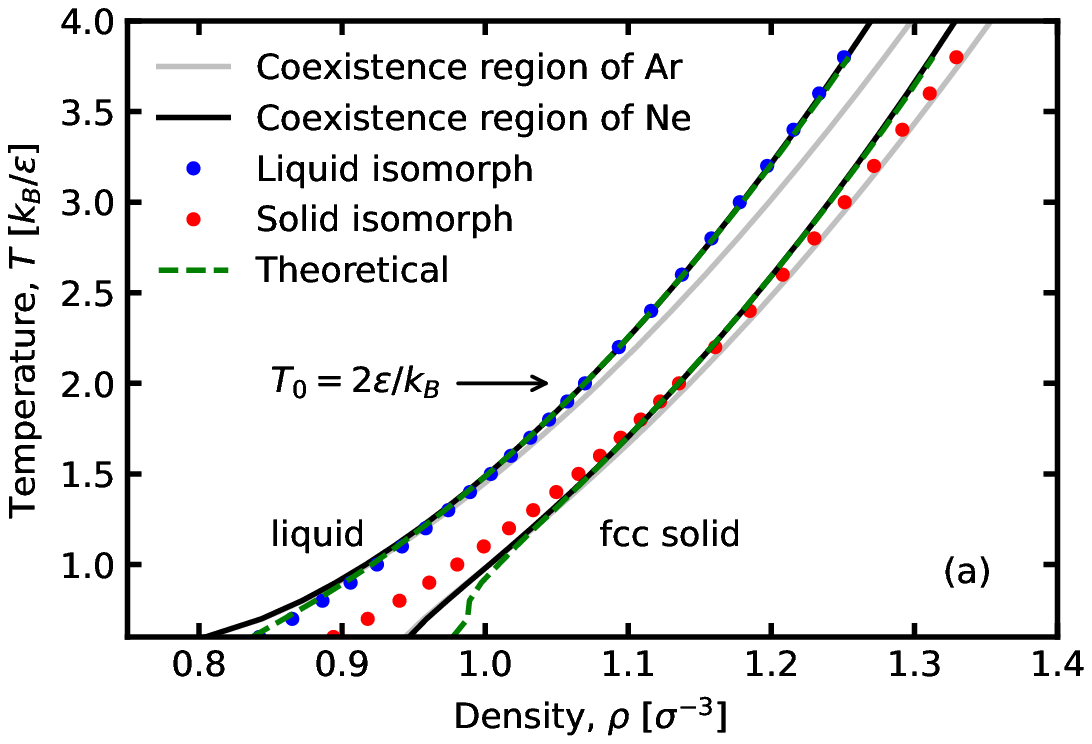}
			\includegraphics[width=0.49\textwidth]{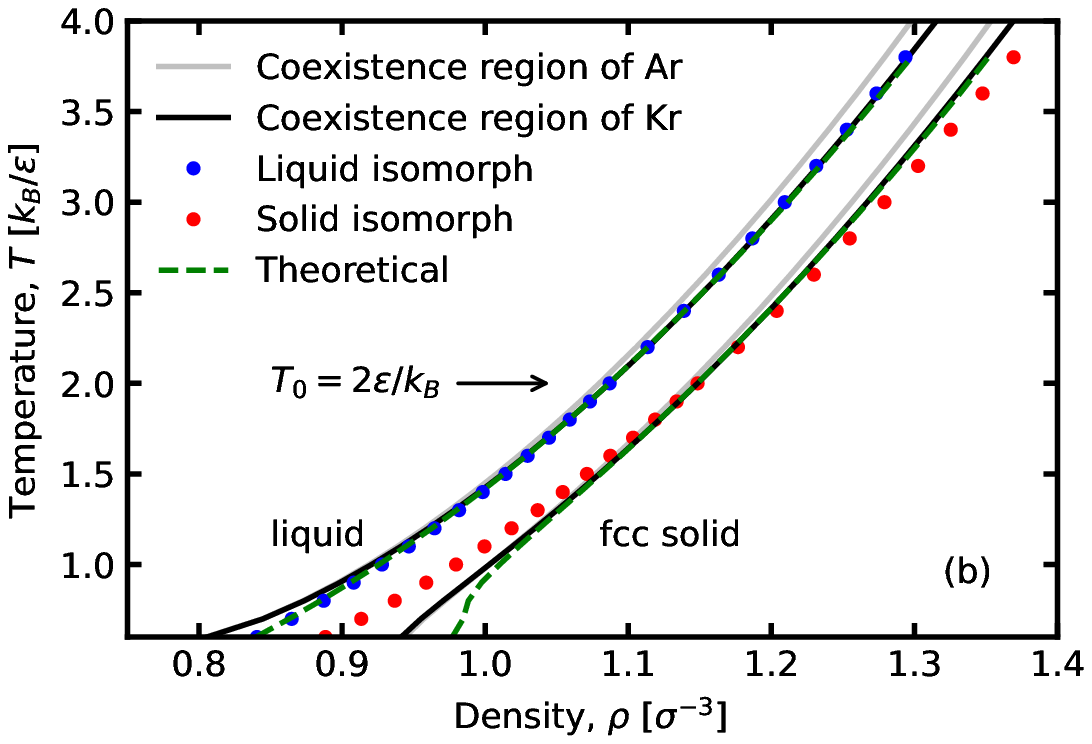}
			\includegraphics[width=0.49\textwidth]{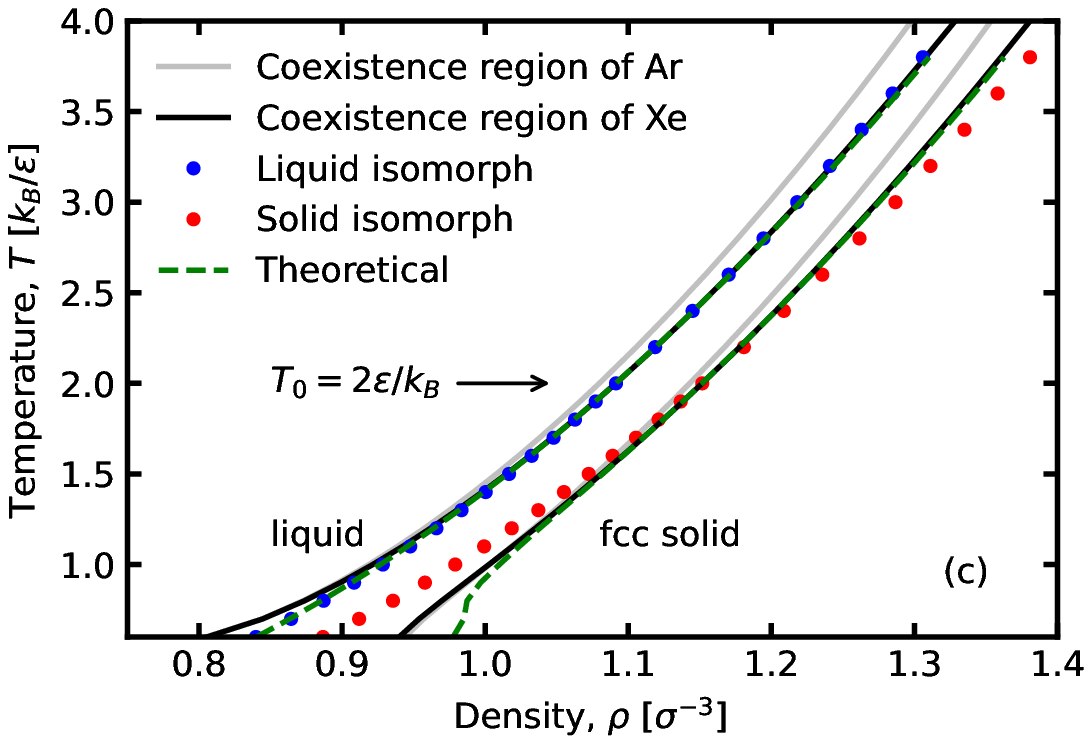}
			\includegraphics[width=0.49\textwidth]{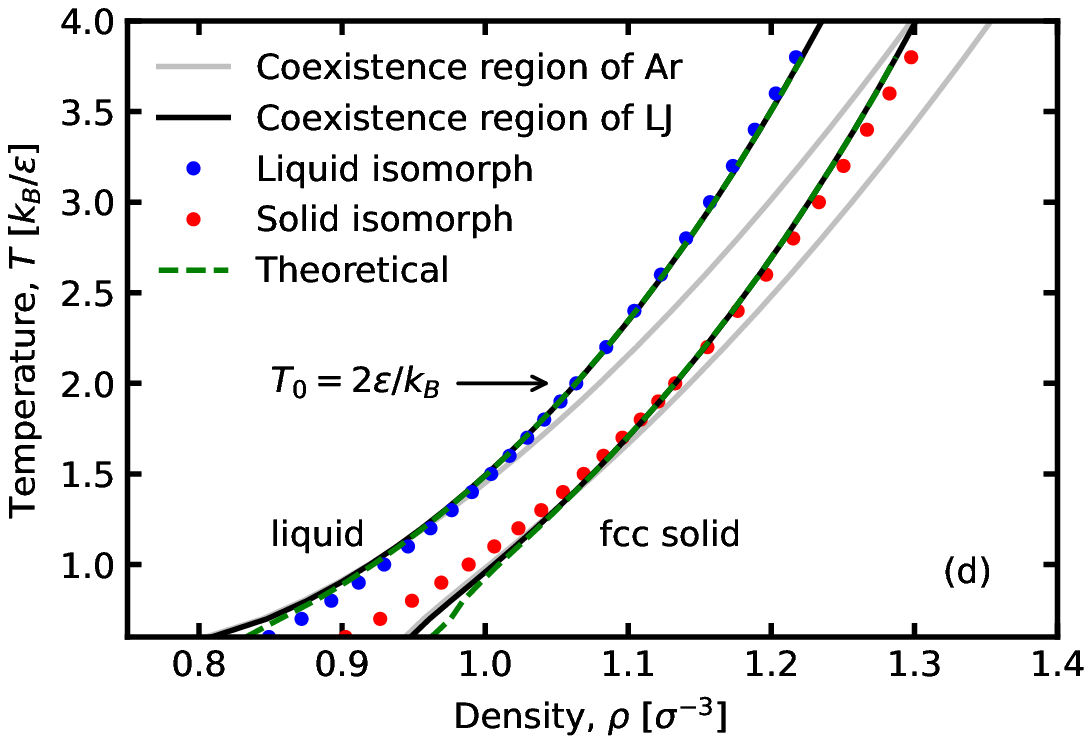}
		\end{center}
		\caption{\label{fig:density_temperature} The solid-liquid coexistence region in the density-temperature plane of (a) neon, (b) krypton, (c) xenon and (d) the LJ model. The gray lines mark the coexistence region of argon. The dots are solid and liquid isomorphic state points of the reference state point at $T_0=2\varepsilon/k_B$, the green dashed line is the theoretical prediction of the isomorph theory \cite{pedersen2016}. The prediction is striking at high densities, but deviates more at low densities near the triple point.}
	\end{figure*}

	Hidden scale invariance implies the existence of lines in the phase diagram along which structure, dynamics and some thermodynamic properties are invariant in given reduced units to a good approximation. These lines, referred to as ``isomorphs'', are configurational adiabats, i.e., lines of constant excess entropy. They can be computed by numerical integration in the logarithmic density-temperature plane of the ``density-scaling exponent'' $\gamma \equiv \left.\frac{\partial \ln T}{\partial \ln \rho}\right|_{S_\textrm{ex}}$. The density-scaling exponent can be computed in an $NVT$ simulation from the fluctuations in potential energy and virial: $\gamma = \langle\Delta W\Delta U\rangle/\langle(\Delta U)^2\rangle
	$ \cite{gnan2009}. Isomorphic state points can then be found by numerical integration, using for instance the recently introduced fourth-order standard Runge-Kutta method \cite{attia2020}. The initial state point for the integration is chosen as the coexistence points at the reference temperature $T_0=2\varepsilon/k_B$ (Table \ref{table:interface_pinning}). The dots on Fig.\ \ref{fig:density_temperature} show the isomorphs in the density-temperature plane of solid (red) and liquid (blue) states for the four noble elements under study. For all the elements, including Ar \cite{paperI}, the isomorphs follow the boundary of the coexistence region (solid lines) with minor deviations. The largest deviations are consistently found for the solid phase near the triple point.  Figure \ref{fig:pressure_temperature} show the same information in the temperature-pressure plane. 
	
\begin{figure}
		\begin{center}
			\includegraphics[width=0.49\textwidth]{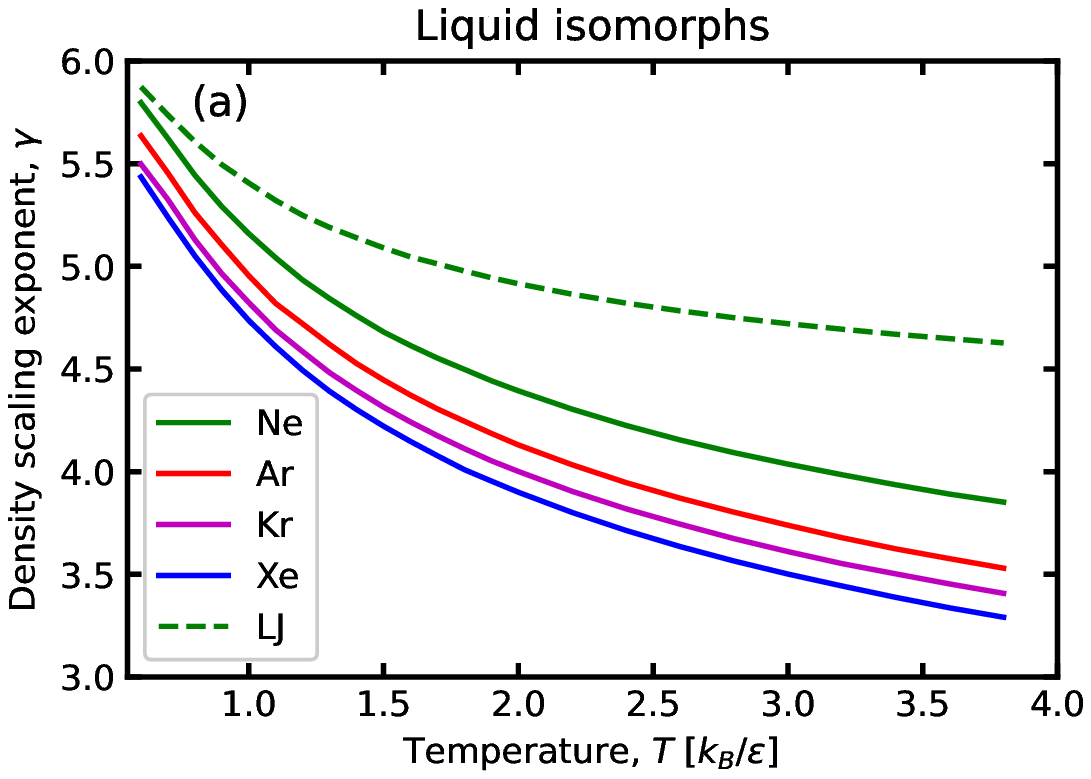}
			\includegraphics[width=0.49\textwidth]{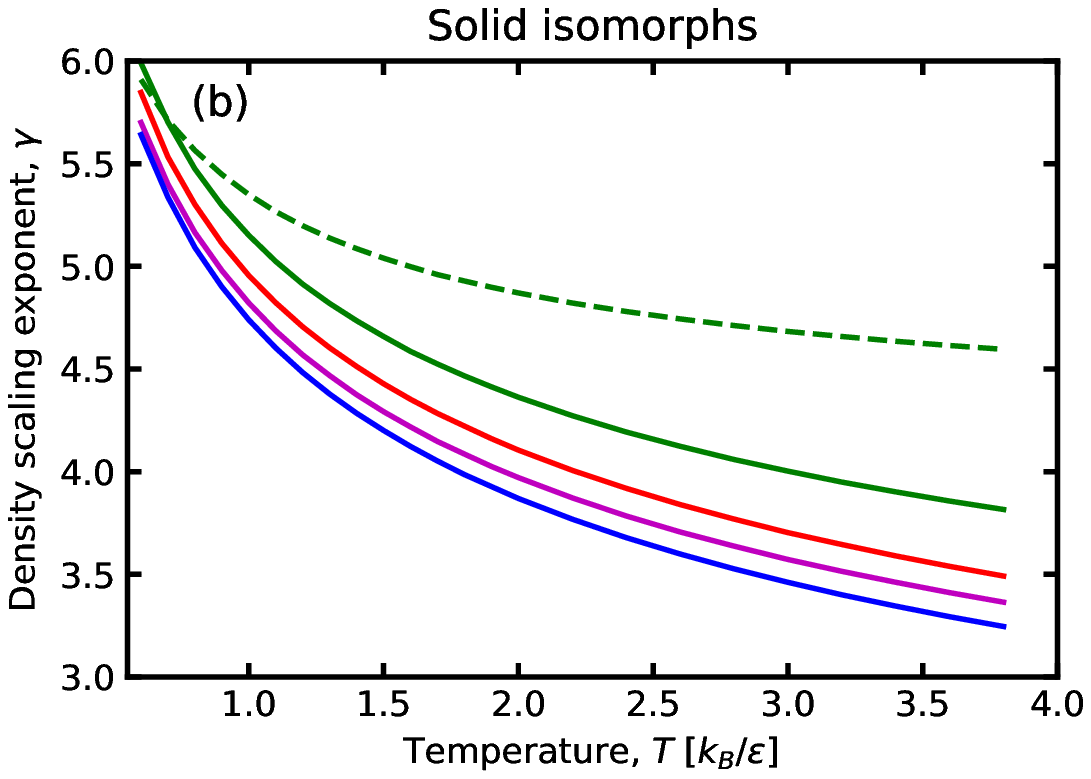}
		\end{center}
		\caption{\label{fig:density_scaling_exponent} The density-scaling exponent $\gamma$ of the liquid isomorphs (a) and the solid isomorphs (b) for the SAAP elements Ne (solid green), Ar (solid red), Kr (solid violet), and LJ (dashed green). }
	\end{figure}

	Figure \ref{fig:density_scaling_exponent}(a) shows the density-scaling exponents of the elements along the liquid isomorphs as a function of the temperature. At low temperatures, near the triple point, the exponents are 5.6$\pm0.3$. This value is close to that of the LJ potential \cite{pedersen2008}. For the SAAP potential, $\gamma$ decrease at higher temperatures and densities as it does for the LJ model; however, the $\gamma$ variation is larger for the SAAP elements and $\gamma$ even goes below the LJ infinite-temperature limit of four. We conclude that the LJ potential is insufficient in describing the configurational adiabats of the noble elements.
	The $\gamma$'s decrease with increasing atomic number. The value of $\gamma$ can be estimated from the pair potential \cite{paperI}, and the decrease of $\gamma$ with increasing atom number is directly related to the softness of the pair interaction: the softer pair interactions of Xe explain why its $\gamma$ is lower than that of Ne. Figure \ref{fig:density_scaling_exponent}(b) shows the density-scaling exponents $\gamma$ of the elements along the solid isomorphs. The conclusions are the same for the liquids.
	
\begin{figure}
		\begin{center}
			\includegraphics[width=0.49\textwidth]{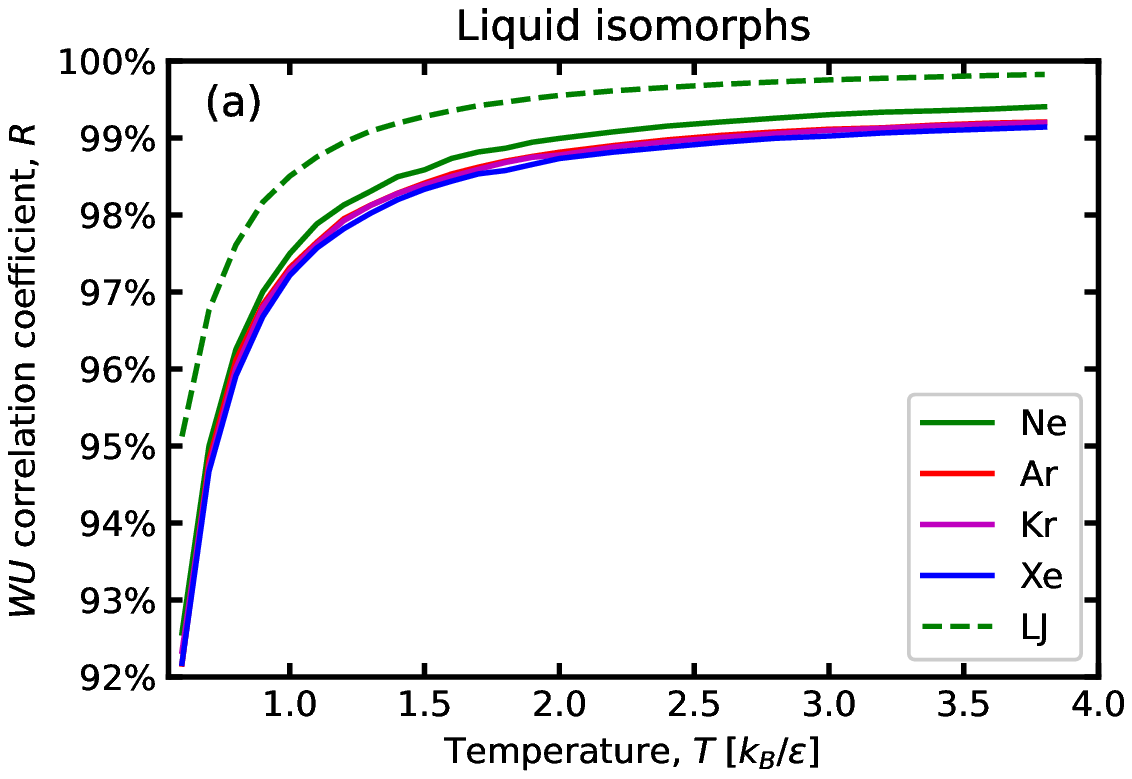}
			\includegraphics[width=0.49\textwidth]{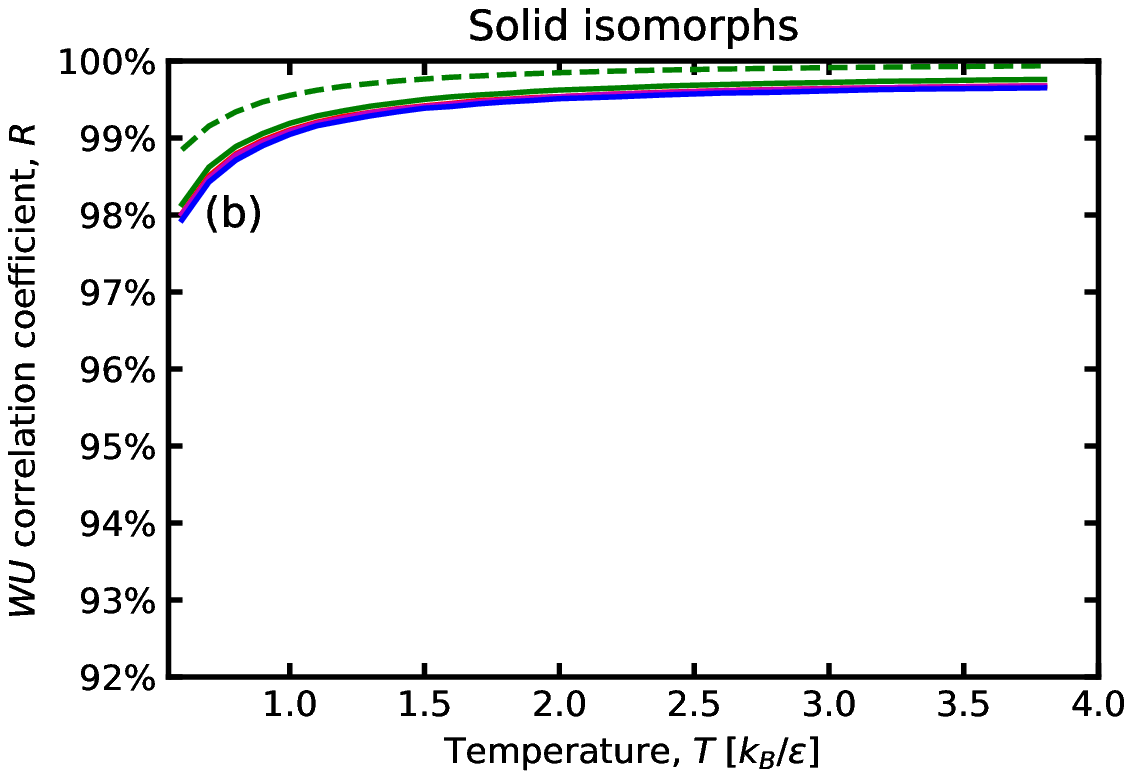}
		\end{center}
		\caption{\label{fig:correlation_coefficient} Person's correlation coefficient $R$ between the virial $W$ and the potential energy $U$. $R$ is computed along the liquid isomorphs (a) and the solid isomorphs (b) and is  shown as a function of temperature $T$. The fact that $R$ is close to unity implies that the potential-energy function has hidden scale invariance along these lines \cite{gnan2009}.}
	\end{figure}

	Figure \ref{fig:correlation_coefficient}(a) shows the Pearson correlation coefficients between the virial  $W$ and the potential energy $U$ in the $NVT$ ensemble defined by  
	$R\equiv\langle\Delta W \Delta U \rangle/\sqrt{\langle(\Delta W)^2\rangle\langle(\Delta U)^2\rangle}.$ The correlation coefficient is close to unity, $R>0.92$, demonstrating that the potential energy function has hidden-scale invariance. Thus the structure, dynamics, and certain thermodynamics quantities are expected to be nearly invariant (in reduced units). This was demonstrated for Ar in Paper I \cite{paperI}. The $WU$-correlation is slightly weaker than for the LJ model (green dashed line on Fig.\ \ref{fig:correlation_coefficient}(a)).
	Figure \ref{fig:correlation_coefficient}(b) shows that $R>0.98$ for the isomorphs of the solid phases.
	
	\begin{figure*}
		\begin{center}
			\includegraphics[width=0.49\textwidth]{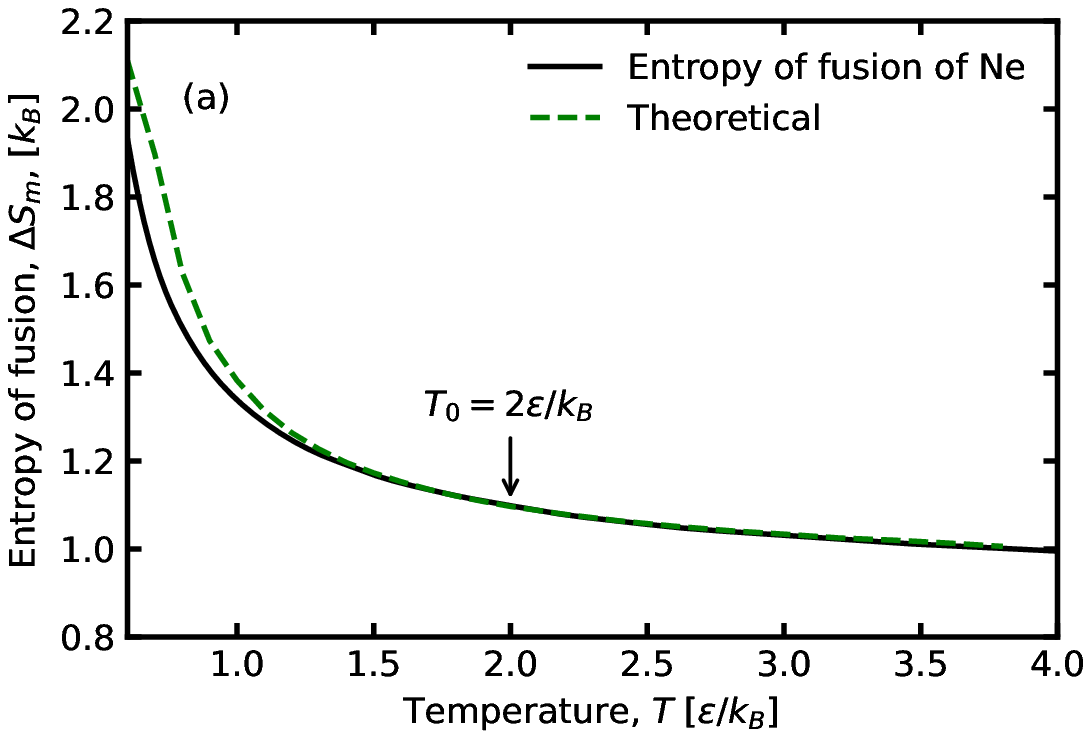}
			\includegraphics[width=0.49\textwidth]{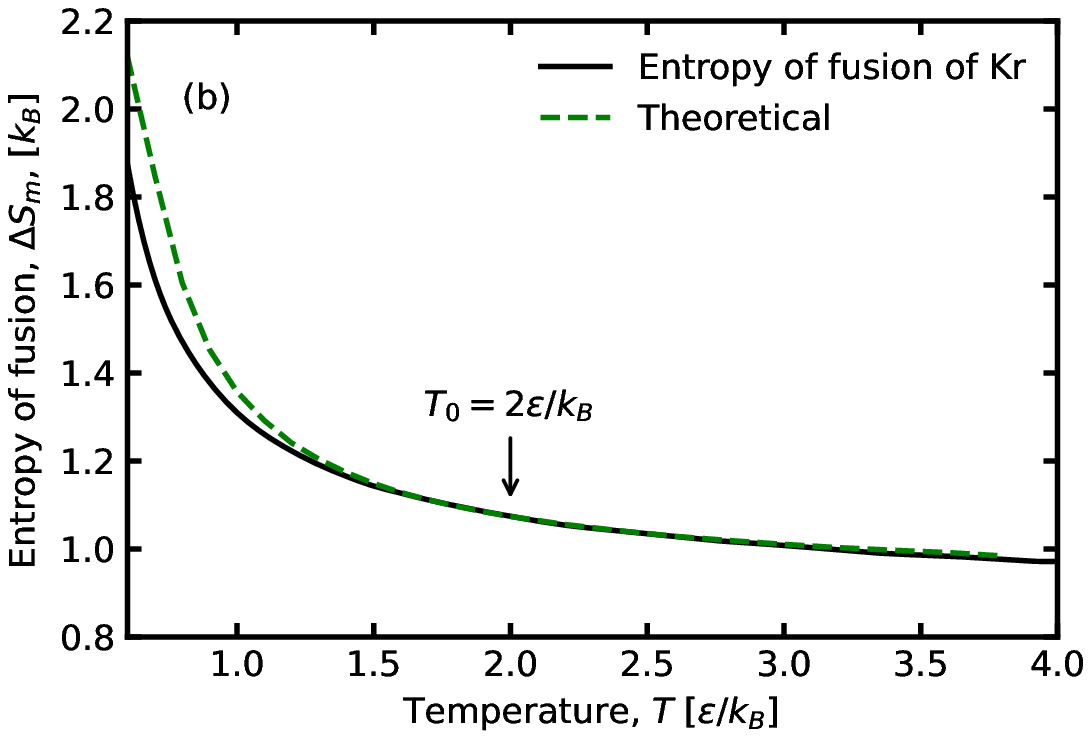}
			\includegraphics[width=0.49\textwidth]{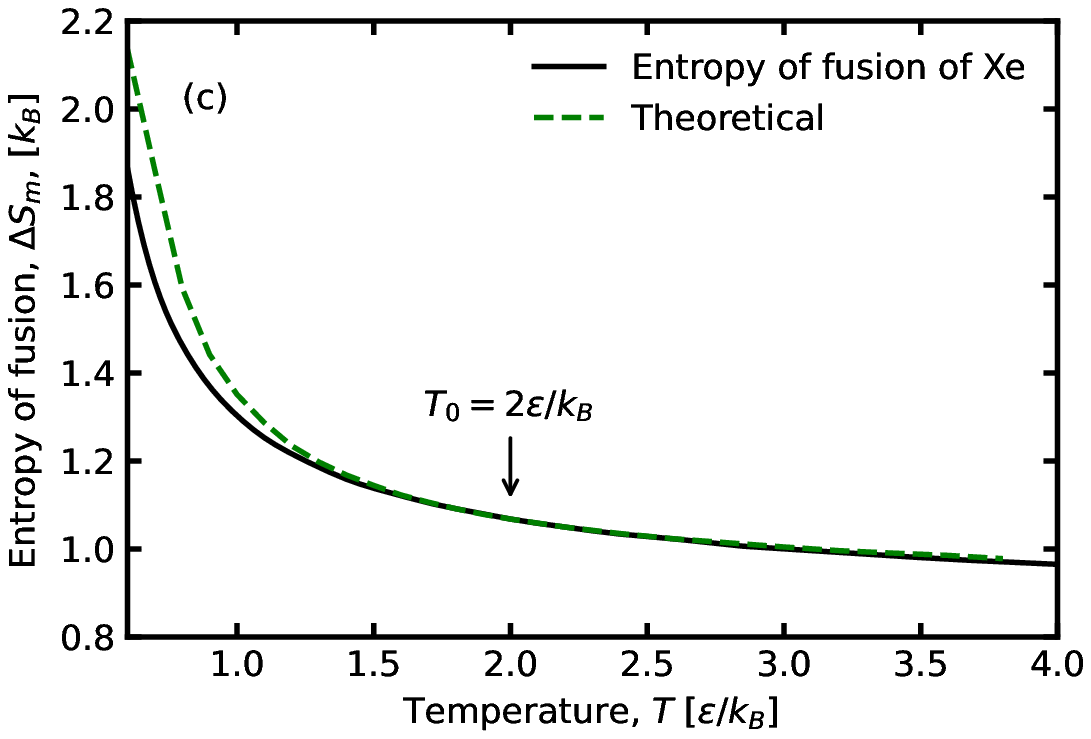}
			\includegraphics[width=0.49\textwidth]{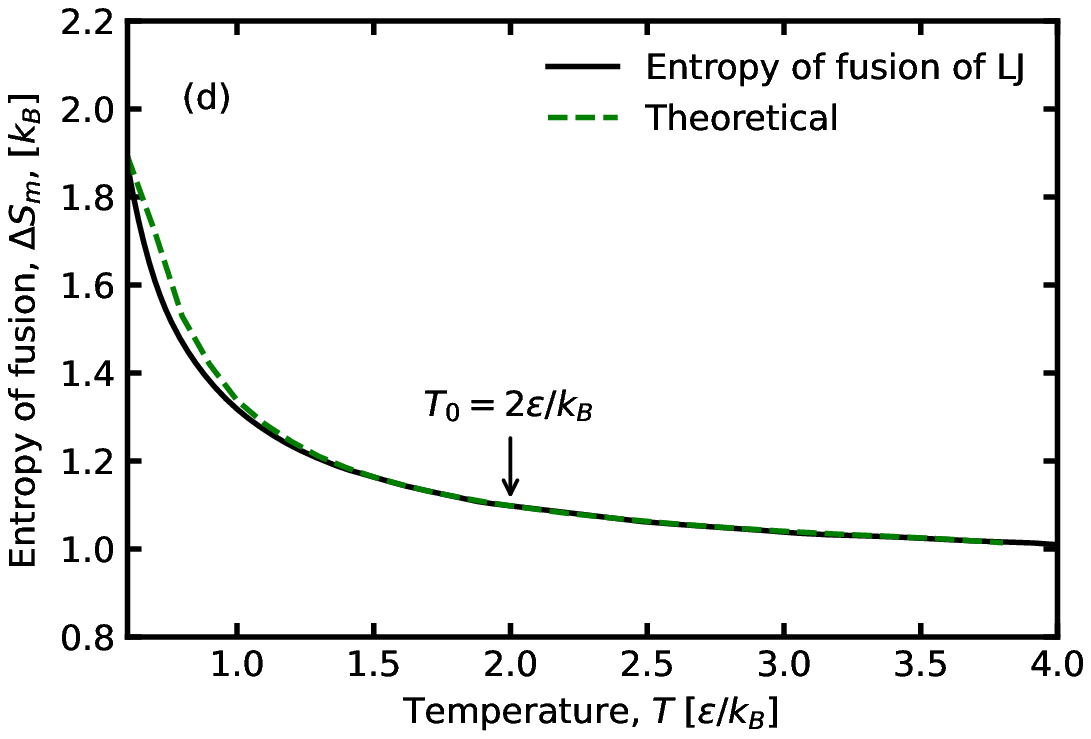}
		\end{center}
		\caption{\label{fig:entropy_of_fusion} The entropy of fusion $\Delta S_m$ of Ne (a), Kr (b), Xe (c) and LJ (d). The solid black lines are the true $\Delta S_m$ and the green dashed lines are the theoretical predictions of the isomorph theory. Hard-sphere based melting theories predict a constant entropy of fusion.}
	\end{figure*}

	\section{Isomorph theory of the solid-liquid coexistence line}
	We have established that the potential-energy functions of the SAAP elements obey hidden scale invariance. This fact allows one to use the framework presented in Reference \cite{pedersen2016} to make theoretical predictions of the shape and property variation along the coexistence line. The basic idea is to make a first-order Taylor expansion from the isothermal state-points along the isomorphs, utilizing the fact that the coexistence lines are almost isomorphs. Property variations along an isomorph can be predicted from a single state point \cite{paperI}, and thus, information is in principle only needed at a single coexistence state point. 
	
	We use information along a liquid and a solid isomorph (computed by numerical integration of $\gamma$).
	The accuracy of the theory was demonstrated for Ar in Paper I \cite{paperI}. The green dashed lines in Figs. \ref{fig:pressure_temperature}(a)-(c) show the theoretical predictions for Ne, Kr, Xe, respectively. The agreement is excellent to the point that the lines are barely visible as they fall on top of the true coexistence lines (shown in solid black). A comparison with the results of the LJ mode (Fig.\ \ref{fig:pressure_temperature}(d)) shows, however, that these results are slightly worse than for the LJ model. This is to be anticipated since the $WU$ correlation coefficient $R$ is weaker (Fig.\ \ref{fig:correlation_coefficient}), and the density scaling exponent $\gamma$ is changing more rapidly (Fig.\ \ref{fig:density_scaling_exponent}) for the SAAP potentials than for the LJ potential. 
	
	Figures \ref{fig:density_temperature}(a)-(c) show the theoretical prediction of the coexistence region's boundaries in the density-temperature plane as green dashed lines. The agreement is sound, but some deviations are noticeable at lower temperatures near the triple point temperature. For these temperatures, the density of the solid isomorphs are several percent lower than the density of melting. These deviations may come from the fact that only the first-order terms in the Taylor expansion were included in the analysis. We hope to investigate this in the future.
	
	The isomorph theory of the coexistence lines predicts both dynamical, structural and thermodynamic properties along the melting line. In Paper I \cite{paperI} this was used for SAAP Ar to predict: i) the value of the diffusion constant along the liquid freezing line, ii) the Lindemann ratio of the solid along the melting line, and iii) the entropy of fusion $\Delta S_m$. As an example, Figs.\ \ref{fig:entropy_of_fusion}(a)-(c) show the latter ($\Delta S_m$) for the remaining elements; Ne, Kr and Xe, respectively. The accuracy is comparable to that of Ar, but again slightly worse than that of the LJ model (Fig. \ref{fig:entropy_of_fusion}(d)).
	For comparison, we note that hard-sphere based melting models predict $\Delta S_m$ to be constant. Thus, the theoretical predictions of the isomorph framework are encouraging.

\section{ACKNOWLEDGMENTS}
The authors thanks Ian Bell, Søren Toxværd, Lorenzo Costigliola, Thomas B.\ Schrøder and Nicholas Bailey for their suggestions during the preparations of this manuscript and support by the VILLUM Foundation’s Matter grant (No. 16515).

\section{SUPPLEMENTARY MATERIAL}
See the supplementary material in Zenodo.org at \url{http://doi.org/10.5281/zenodo.3888373} for the raw data, and additional graphs.

\section{DATA AVAILABILITY}
The data that support the findings of this study are openly available in Zenodo.org at \url{http://doi.org/10.5281/zenodo.3888373}, and the supplementary material.
	
\bibliography{references}

\begin{thebibliography}{18}%
\makeatletter
\providecommand \@ifxundefined [1]{%
 \@ifx{#1\undefined}
}%
\providecommand \@ifnum [1]{%
 \ifnum #1\expandafter \@firstoftwo
 \else \expandafter \@secondoftwo
 \fi
}%
\providecommand \@ifx [1]{%
 \ifx #1\expandafter \@firstoftwo
 \else \expandafter \@secondoftwo
 \fi
}%
\providecommand \natexlab [1]{#1}%
\providecommand \enquote  [1]{``#1''}%
\providecommand \bibnamefont  [1]{#1}%
\providecommand \bibfnamefont [1]{#1}%
\providecommand \citenamefont [1]{#1}%
\providecommand \href@noop [0]{\@secondoftwo}%
\providecommand \href [0]{\begingroup \@sanitize@url \@href}%
\providecommand \@href[1]{\@@startlink{#1}\@@href}%
\providecommand \@@href[1]{\endgroup#1\@@endlink}%
\providecommand \@sanitize@url [0]{\catcode `\\12\catcode `\$12\catcode
  `\&12\catcode `\#12\catcode `\^12\catcode `\_12\catcode `\%12\relax}%
\providecommand \@@startlink[1]{}%
\providecommand \@@endlink[0]{}%
\providecommand \url  [0]{\begingroup\@sanitize@url \@url }%
\providecommand \@url [1]{\endgroup\@href {#1}{\urlprefix }}%
\providecommand \urlprefix  [0]{URL }%
\providecommand \Eprint [0]{\href }%
\providecommand \doibase [0]{http://dx.doi.org/}%
\providecommand \selectlanguage [0]{\@gobble}%
\providecommand \bibinfo  [0]{\@secondoftwo}%
\providecommand \bibfield  [0]{\@secondoftwo}%
\providecommand \translation [1]{[#1]}%
\providecommand \BibitemOpen [0]{}%
\providecommand \bibitemStop [0]{}%
\providecommand \bibitemNoStop [0]{.\EOS\space}%
\providecommand \EOS [0]{\spacefactor3000\relax}%
\providecommand \BibitemShut  [1]{\csname bibitem#1\endcsname}%
\let\auto@bib@innerbib\@empty
\bibitem [{\citenamefont {Singh}\ \emph {et~al.}(2020)\citenamefont {Singh},
  \citenamefont {Dyre},\ and\ \citenamefont {Pedersen}}]{paperI}%
  \BibitemOpen
  \bibfield  {author} {\bibinfo {author} {\bibfnamefont {A.~N.}\ \bibnamefont
  {Singh}}, \bibinfo {author} {\bibfnamefont {J.~C.}\ \bibnamefont {Dyre}}, \
  and\ \bibinfo {author} {\bibfnamefont {U.~R.}\ \bibnamefont {Pedersen}},\
  }\href@noop {} {\  (\bibinfo {year} {2020})}\BibitemShut {NoStop}%
\bibitem [{\citenamefont {Deiters}\ and\ \citenamefont
  {Sadus}(2019{\natexlab{a}})}]{deiters2019}%
  \BibitemOpen
  \bibfield  {author} {\bibinfo {author} {\bibfnamefont {U.~K.}\ \bibnamefont
  {Deiters}}\ and\ \bibinfo {author} {\bibfnamefont {R.~J.}\ \bibnamefont
  {Sadus}},\ }\href {\doibase 10.1063/1.5085420} {\bibfield  {journal}
  {\bibinfo  {journal} {J. Chem. Phys.}\ }\textbf {\bibinfo {volume} {150}},\
  \bibinfo {pages} {134504} (\bibinfo {year} {2019}{\natexlab{a}})}\BibitemShut
  {NoStop}%
\bibitem [{\citenamefont {Patkowski}\ and\ \citenamefont
  {Szalewicz}(2010)}]{konrad2010}%
  \BibitemOpen
  \bibfield  {author} {\bibinfo {author} {\bibfnamefont {K.}~\bibnamefont
  {Patkowski}}\ and\ \bibinfo {author} {\bibfnamefont {K.}~\bibnamefont
  {Szalewicz}},\ }\href {\doibase 10.1063/1.3478513} {\bibfield  {journal}
  {\bibinfo  {journal} {The Journal of Chemical Physics}\ }\textbf {\bibinfo
  {volume} {133}},\ \bibinfo {pages} {094304} (\bibinfo {year}
  {2010})}\BibitemShut {NoStop}%
\bibitem [{\citenamefont {Bartlett}\ and\ \citenamefont
  {Musia\l{}}(2007)}]{bartlett2007}%
  \BibitemOpen
  \bibfield  {author} {\bibinfo {author} {\bibfnamefont {R.~J.}\ \bibnamefont
  {Bartlett}}\ and\ \bibinfo {author} {\bibfnamefont {M.}~\bibnamefont
  {Musia\l{}}},\ }\href {\doibase 10.1103/RevModPhys.79.291} {\bibfield
  {journal} {\bibinfo  {journal} {Rev. Mod. Phys.}\ }\textbf {\bibinfo {volume}
  {79}},\ \bibinfo {pages} {291} (\bibinfo {year} {2007})}\BibitemShut
  {NoStop}%
\bibitem [{\citenamefont {Nasrabad}\ \emph {et~al.}(2004)\citenamefont
  {Nasrabad}, \citenamefont {Laghaei},\ and\ \citenamefont
  {Deiters}}]{nasrabad2004}%
  \BibitemOpen
  \bibfield  {author} {\bibinfo {author} {\bibfnamefont {A.~E.}\ \bibnamefont
  {Nasrabad}}, \bibinfo {author} {\bibfnamefont {R.}~\bibnamefont {Laghaei}}, \
  and\ \bibinfo {author} {\bibfnamefont {U.~K.}\ \bibnamefont {Deiters}},\
  }\href {\doibase 10.1063/1.1783271} {\bibfield  {journal} {\bibinfo
  {journal} {J. Chem. Phys.}\ }\textbf {\bibinfo {volume} {121}},\ \bibinfo
  {pages} {6423} (\bibinfo {year} {2004})}\BibitemShut {NoStop}%
\bibitem [{\citenamefont {Smits}\ \emph {et~al.}(2020)\citenamefont {Smits},
  \citenamefont {Jerabek}, \citenamefont {Pahl},\ and\ \citenamefont
  {Schwerdtfeger}}]{smits2020}%
  \BibitemOpen
  \bibfield  {author} {\bibinfo {author} {\bibfnamefont {O.~R.}\ \bibnamefont
  {Smits}}, \bibinfo {author} {\bibfnamefont {P.}~\bibnamefont {Jerabek}},
  \bibinfo {author} {\bibfnamefont {E.}~\bibnamefont {Pahl}}, \ and\ \bibinfo
  {author} {\bibfnamefont {P.}~\bibnamefont {Schwerdtfeger}},\ }\href {\doibase
  10.1103/PhysRevB.101.104103} {\bibfield  {journal} {\bibinfo  {journal}
  {Phys. Rev. B}\ }\textbf {\bibinfo {volume} {101}},\ \bibinfo {pages}
  {104103} (\bibinfo {year} {2020})}\BibitemShut {NoStop}%
\bibitem [{\citenamefont {Bailey}\ \emph {et~al.}(2017)\citenamefont {Bailey},
  \citenamefont {Ingebrigtsen}, \citenamefont {Hansen}, \citenamefont
  {Veldhorst}, \citenamefont {Bøhling}, \citenamefont {Lemarchand},
  \citenamefont {Olsen}, \citenamefont {Bacher}, \citenamefont {Costigliola},
  \citenamefont {Pedersen}, \citenamefont {Larsen}, \citenamefont {Dyre},\ and\
  \citenamefont {Schrøder}}]{rumd}%
  \BibitemOpen
  \bibfield  {author} {\bibinfo {author} {\bibfnamefont {N.~P.}\ \bibnamefont
  {Bailey}}, \bibinfo {author} {\bibfnamefont {T.~S.}\ \bibnamefont
  {Ingebrigtsen}}, \bibinfo {author} {\bibfnamefont {J.~S.}\ \bibnamefont
  {Hansen}}, \bibinfo {author} {\bibfnamefont {A.~A.}\ \bibnamefont
  {Veldhorst}}, \bibinfo {author} {\bibfnamefont {L.}~\bibnamefont {Bøhling}},
  \bibinfo {author} {\bibfnamefont {C.~A.}\ \bibnamefont {Lemarchand}},
  \bibinfo {author} {\bibfnamefont {A.~E.}\ \bibnamefont {Olsen}}, \bibinfo
  {author} {\bibfnamefont {A.~K.}\ \bibnamefont {Bacher}}, \bibinfo {author}
  {\bibfnamefont {L.}~\bibnamefont {Costigliola}}, \bibinfo {author}
  {\bibfnamefont {U.~R.}\ \bibnamefont {Pedersen}}, \bibinfo {author}
  {\bibfnamefont {H.}~\bibnamefont {Larsen}}, \bibinfo {author} {\bibfnamefont
  {J.~C.}\ \bibnamefont {Dyre}}, \ and\ \bibinfo {author} {\bibfnamefont
  {T.~B.}\ \bibnamefont {Schrøder}},\ }\href {\doibase
  10.21468/SciPostPhys.3.6.038} {\bibfield  {journal} {\bibinfo  {journal}
  {SciPost Phys.}\ }\textbf {\bibinfo {volume} {3}},\ \bibinfo {pages} {038}
  (\bibinfo {year} {2017})}\BibitemShut {NoStop}%
\bibitem [{\citenamefont {Grønbech-Jensen}\ \emph {et~al.}(2014)\citenamefont
  {Grønbech-Jensen}, \citenamefont {Hayre},\ and\ \citenamefont
  {Farago}}]{gronbech2014}%
  \BibitemOpen
  \bibfield  {author} {\bibinfo {author} {\bibfnamefont {N.}~\bibnamefont
  {Grønbech-Jensen}}, \bibinfo {author} {\bibfnamefont {N.~R.}\ \bibnamefont
  {Hayre}}, \ and\ \bibinfo {author} {\bibfnamefont {O.}~\bibnamefont
  {Farago}},\ }\href {\doibase 10.1016/j.cpc.2013.10.006} {\bibfield  {journal}
  {\bibinfo  {journal} {Comput. Phys. Commun.}\ }\textbf {\bibinfo {volume}
  {185}},\ \bibinfo {pages} {524 } (\bibinfo {year} {2014})}\BibitemShut
  {NoStop}%
\bibitem [{\citenamefont {Pedersen}(2013)}]{pedersen2013}%
  \BibitemOpen
  \bibfield  {author} {\bibinfo {author} {\bibfnamefont {U.~R.}\ \bibnamefont
  {Pedersen}},\ }\href {\doibase 10.1063/1.4818747} {\bibfield  {journal}
  {\bibinfo  {journal} {J. Chem. Phys.}\ }\textbf {\bibinfo {volume} {139}},\
  \bibinfo {pages} {104102} (\bibinfo {year} {2013})}\BibitemShut {NoStop}%
\bibitem [{\citenamefont {Vos}\ \emph {et~al.}(1991)\citenamefont {Vos},
  \citenamefont {Schouten}, \citenamefont {Young},\ and\ \citenamefont
  {Ross}}]{vos1991}%
  \BibitemOpen
  \bibfield  {author} {\bibinfo {author} {\bibfnamefont {W.~L.}\ \bibnamefont
  {Vos}}, \bibinfo {author} {\bibfnamefont {J.~A.}\ \bibnamefont {Schouten}},
  \bibinfo {author} {\bibfnamefont {D.~A.}\ \bibnamefont {Young}}, \ and\
  \bibinfo {author} {\bibfnamefont {M.}~\bibnamefont {Ross}},\ }\href {\doibase
  10.1063/1.460683} {\bibfield  {journal} {\bibinfo  {journal} {J. Chem.
  Phys.}\ }\textbf {\bibinfo {volume} {94}},\ \bibinfo {pages} {3835} (\bibinfo
  {year} {1991})},\ \Eprint
  {http://arxiv.org/abs/https://doi.org/10.1063/1.460683}
  {https://doi.org/10.1063/1.460683} \BibitemShut {NoStop}%
\bibitem [{\citenamefont {Ferreira}\ and\ \citenamefont
  {Lobo}(2008)}]{ferreira2008}%
  \BibitemOpen
  \bibfield  {author} {\bibinfo {author} {\bibfnamefont {A.}~\bibnamefont
  {Ferreira}}\ and\ \bibinfo {author} {\bibfnamefont {L.}~\bibnamefont
  {Lobo}},\ }\href {\doibase https://doi.org/10.1016/j.jct.2007.11.007}
  {\bibfield  {journal} {\bibinfo  {journal} {J. Chem. Thermodyn.}\ }\textbf
  {\bibinfo {volume} {40}},\ \bibinfo {pages} {618 } (\bibinfo {year}
  {2008})}\BibitemShut {NoStop}%
\bibitem [{\citenamefont {Deiters}\ and\ \citenamefont
  {Sadus}(2019{\natexlab{b}})}]{deiters2019b}%
  \BibitemOpen
  \bibfield  {author} {\bibinfo {author} {\bibfnamefont {U.~K.}\ \bibnamefont
  {Deiters}}\ and\ \bibinfo {author} {\bibfnamefont {R.~J.}\ \bibnamefont
  {Sadus}},\ }\href {\doibase 10.1063/1.5109052} {\bibfield  {journal}
  {\bibinfo  {journal} {J. Chem. Phys.}\ }\textbf {\bibinfo {volume} {151}},\
  \bibinfo {pages} {034509} (\bibinfo {year} {2019}{\natexlab{b}})}\BibitemShut
  {NoStop}%
\bibitem [{\citenamefont {Mastny}\ and\ \citenamefont
  {de~Pablo}(2007)}]{mastny2007}%
  \BibitemOpen
  \bibfield  {author} {\bibinfo {author} {\bibfnamefont {E.~A.}\ \bibnamefont
  {Mastny}}\ and\ \bibinfo {author} {\bibfnamefont {J.~J.}\ \bibnamefont
  {de~Pablo}},\ }\href {\doibase 10.1063/1.2753149} {\bibfield  {journal}
  {\bibinfo  {journal} {J. Chem. Phys}\ }\textbf {\bibinfo {volume} {127}},\
  \bibinfo {pages} {104504} (\bibinfo {year} {2007})}\BibitemShut {NoStop}%
\bibitem [{\citenamefont {Simon F.~E.}(1929)}]{simon1929}%
  \BibitemOpen
  \bibfield  {author} {\bibinfo {author} {\bibfnamefont {Z.}~\bibnamefont
  {Simon F.~E.}, \bibfnamefont {Glatzel~G.}},\ }\href {\doibase
  10.1002/zaac.19291780123} {\bibfield  {journal} {\bibinfo  {journal} {Anorg.
  (Allg.) Chem.}\ }\textbf {\bibinfo {volume} {178}},\ \bibinfo {pages}
  {309–312} (\bibinfo {year} {1929})}\BibitemShut {NoStop}%
\bibitem [{\citenamefont {Gnan}\ \emph {et~al.}(2009)\citenamefont {Gnan},
  \citenamefont {Schrøder}, \citenamefont {Pedersen}, \citenamefont {Bailey},\
  and\ \citenamefont {Dyre}}]{gnan2009}%
  \BibitemOpen
  \bibfield  {author} {\bibinfo {author} {\bibfnamefont {N.}~\bibnamefont
  {Gnan}}, \bibinfo {author} {\bibfnamefont {T.~B.}\ \bibnamefont {Schrøder}},
  \bibinfo {author} {\bibfnamefont {U.~R.}\ \bibnamefont {Pedersen}}, \bibinfo
  {author} {\bibfnamefont {N.~P.}\ \bibnamefont {Bailey}}, \ and\ \bibinfo
  {author} {\bibfnamefont {J.~C.}\ \bibnamefont {Dyre}},\ }\href {\doibase
  10.1063/1.3265957} {\bibfield  {journal} {\bibinfo  {journal} {J. Chem.
  Phys.}\ }\textbf {\bibinfo {volume} {131}},\ \bibinfo {pages} {234504}
  (\bibinfo {year} {2009})}\BibitemShut {NoStop}%
\bibitem [{\citenamefont {Attia}\ \emph {et~al.}(2020)\citenamefont {Attia},
  \citenamefont {Dyre},\ and\ \citenamefont {Pedersen}}]{attia2020}%
  \BibitemOpen
  \bibfield  {author} {\bibinfo {author} {\bibfnamefont {E.}~\bibnamefont
  {Attia}}, \bibinfo {author} {\bibfnamefont {J.~C.}\ \bibnamefont {Dyre}}, \
  and\ \bibinfo {author} {\bibfnamefont {U.~R.}\ \bibnamefont {Pedersen}},\
  }\href@noop {} {\bibfield  {journal} {\bibinfo  {journal} {{\it preprint}}\ }
  (\bibinfo {year} {2020})}\BibitemShut {NoStop}%
\bibitem [{\citenamefont {Pedersen}\ \emph {et~al.}(2008)\citenamefont
  {Pedersen}, \citenamefont {Bailey}, \citenamefont {Schr\o{}der},\ and\
  \citenamefont {Dyre}}]{pedersen2008}%
  \BibitemOpen
  \bibfield  {author} {\bibinfo {author} {\bibfnamefont {U.~R.}\ \bibnamefont
  {Pedersen}}, \bibinfo {author} {\bibfnamefont {N.~P.}\ \bibnamefont
  {Bailey}}, \bibinfo {author} {\bibfnamefont {T.~B.}\ \bibnamefont
  {Schr\o{}der}}, \ and\ \bibinfo {author} {\bibfnamefont {J.~C.}\ \bibnamefont
  {Dyre}},\ }\href {\doibase 10.1103/PhysRevLett.100.015701} {\bibfield
  {journal} {\bibinfo  {journal} {Phys. Rev. Lett.}\ }\textbf {\bibinfo
  {volume} {100}},\ \bibinfo {pages} {015701} (\bibinfo {year}
  {2008})}\BibitemShut {NoStop}%
\bibitem [{\citenamefont {Pedersen}\ \emph {et~al.}(2016)\citenamefont
  {Pedersen}, \citenamefont {Costigliola}, \citenamefont {Bailey},
  \citenamefont {Schrøder},\ and\ \citenamefont {Dyre}}]{pedersen2016}%
  \BibitemOpen
  \bibfield  {author} {\bibinfo {author} {\bibfnamefont {U.~R.}\ \bibnamefont
  {Pedersen}}, \bibinfo {author} {\bibfnamefont {L.}~\bibnamefont
  {Costigliola}}, \bibinfo {author} {\bibfnamefont {N.~P.}\ \bibnamefont
  {Bailey}}, \bibinfo {author} {\bibfnamefont {T.~B.}\ \bibnamefont
  {Schrøder}}, \ and\ \bibinfo {author} {\bibfnamefont {J.~C.}\ \bibnamefont
  {Dyre}},\ }\href {\doibase 10.1038/ncomms12386} {\bibfield  {journal}
  {\bibinfo  {journal} {Nat. Commun.}\ }\textbf {\bibinfo {volume} {7}},\
  \bibinfo {pages} {12386} (\bibinfo {year} {2016})}\BibitemShut {NoStop}%
\end{thebibliography}%

\end{document}